\documentclass[aps,floatfix,twocolumn]{revtex4}
\usepackage{hyperref}
\usepackage{color}
\usepackage[dvipsnames]{xcolor}
\usepackage{amssymb}
\usepackage{amsmath}
\usepackage{bm}
\usepackage{graphicx}
\usepackage{siunitx}
\usepackage{braket}
\usepackage[final]{microtype}

\newcommand{\ee}{\mathrm{e}}
\newcommand\diff{\mathop{}\!d}

\begin{document}

\title{Anomalous Lifetimes of Ultracold Complexes Decaying into a Single Channel: \\
What's Taking So Long in There?}

\author{James F. E. Croft}
\affiliation{The Dodd-Walls Centre for Photonic and Quantum Technologies, New Zealand}
\affiliation{Department of Physics, University of Otago, Dunedin, New Zealand}
\author{John L. Bohn}
\affiliation{JILA, NIST, and Department of Physics, University of Colorado, Boulder, Colorado 80309-0440, USA}
\author{Goulven Qu{\'e}m{\'e}ner}
\affiliation{Universit\'{e} Paris-Saclay, CNRS, Laboratoire Aim\'{e} Cotton, 91405 Orsay, France}

\begin{abstract}
  We investigate the lifetimes of complexes formed in ultracold molecule
  collisions.
  Employing both transition-state-theory and an optical model approach we examine processes that can extend the lifetime of complexes beyond that
  predicted by Rice-Ramsperger-Kassel-Marcus theory.
  We focus on complexes that possess  only one open channel, and find that the extreme
  distribution of widths for this case favors  low decay rates.  Thus decay from a complex into a single energetically available channel can  be anomalously slow, and moreover nonexponential in time.
  We apply the theory to two systems of current experimental interest, RbCs and NaRb, finding qualitatively that the empirical time scales can be accounted for in the theory.
  \end{abstract}

\maketitle

\section{Introduction}
Ultracold molecules live at the fascinating intersection of simplicity and
complexity.
While molecules can be produced in their absolute ground
state~\cite{danzl.haller.ea:quantum,ni.ospelkaus.ea:high} and their long-range collisional dynamics
described by a single partial wave, their short-range collisional dynamics is
highly intricate~\cite{croft.makrides.ea:universality}.
It is this combination of simplicity and complexity which makes ultracold
molecules such a powerful tool for the study of fundamental collisional
mechanisms~\cite{bohn.rey.ea:cold}.

However there is currently a specter haunting the field, the specter of sticky collisions.  Diatomic molecules are
observed to collide and \emph{go away}, as if perishing in a chemical reaction,
even those that are not chemically
reactive at zero temperature~\cite{takekoshi.reichsollner.ea:ultracold,park.will.ea:ultracold,
ye.guo.ea:collisions,guo.ye.ea:dipolar,yang.zhang.ea:observation,
gregory.frye.ea:sticky,gregory.blackmore.ea:loss,gersema.voges.ea:probing,
bause.schindewolf.ea:collisions,he.ye.ea:observation}.
Rather, it is believed that these diatomic molecules enter into a holding pattern,
a four-atom collision complex of unprecedented duration,
where they weave their chaotic tales in secret,
before once more emerging into the daylight of the laboratory.
They have been observed to dwell inside these complexes for extraordinarily long
time scales approaching milliseconds~\cite{gregory.blackmore.ea:loss,
gersema.voges.ea:probing,bause.schindewolf.ea:collisions}.
In at least one case, the same millisecond time scale holds for molecules
colliding with individual atoms~\cite{nichols.liu.ea:detection}.
Such collisions are colloquially referred to as ``sticky'', following the
terminology of Bethe from nuclear physics~\cite{bethe:continuum}.

The \emph{idea} of long-lived, sticky collision complexes is an appealing one.
Owing to their deep, barrierless potential energy surfaces and rather large masses,
four-body complexes of alkali atoms might be expected to possess a high density
of states $\rho$ in which the atoms are unable to dissociate and can get
randomly stuck.
The added feature of ultracold temperatures ensures that molecules which
originate in their ground states can only dissociate into a single open
channel $N_\mathrm{o}=1$.
In this circumstance, it might be expected that the lifetime of the complex is
proportional to $\rho$ and inversely proportional to $N_\mathrm{o}$,
expressed in the simplest way by the Rice-Ramsperger-Kassel-Marcus (RRKM)
expression~\cite{levine:molecular}
\begin{equation}\label{eq:RRKM_lifetime}
\tau \approx \tau_\mathrm{RRKM} = \frac{2\pi\hbar\rho}{N_\mathrm{o}}.
\end{equation}
Considerations such as these led to early speculations of sticking in
ultracold molecules~\cite{Bohn02_PRL,mayle.ruzic.ea:statistical,
mayle.quemener.ea:scattering}.

This naive expectation however proves inadequate.
High-quality estimates of the relevant densities of states reveal that realistic
RRKM lifetimes are shorter than those observed. The longest predicted lifetime is \SI{0.25}{\milli\second} for the
heaviest species RbCs~\cite{christianen.karman.ea:quasiclassical}.
While this particular estimate does nearly agree with the measured lifetime of
(RbCs)$_2^*$~\cite{gregory.blackmore.ea:loss}, other experiments that have
measured the lifetime find results orders of magnitude larger than the RRKM
lifetime~\cite{bause.schindewolf.ea:collisions,gersema.voges.ea:probing,
nichols.liu.ea:detection}.
A caveat is relevant here: we are discussing the lifetimes of the complex
\emph{in the dark}, that is, in transient intervals where light from the
optical dipole trap that confines the molecules is temporarily turned off.
It is established that the complex's lifetimes are significantly reduced by
scattering photons of the trapping light~\cite{christianen.zwierlein.ea:photoinduced,
liu.hu.ea:photo-excitation}, but that is not our concern here.

We are instead interested in the natural, light-free lifetimes of the complexes,
and how they can come to be  far longer than the RRKM lifetimes.
To understand this situation, we consider the time-dependence of a complex assumed
to consist of many resonances, each with its own characteristic decay rate,
following an approach by Peskin \emph{et al}~\cite{peskin.reisler.ea:on}.
This approach gives an effective decay rate that agrees with the RRKM rate in
certain limits, yet can deviate under other circumstances.
In particular, we find that in the limit of a very small number of open channels,
especially the $N_\mathrm{o}=1$ case of particular interest to ultracold molecule
applications, the decay of the complexes can be significantly slower than the
RRKM result.
In this context, we note that for decay of the chemically reactive species KRb,
the collision complexes do appear to decay on timescales consistent with the
RRKM result~\cite{liu.hu.ea:photo-excitation}, a comforting fact, since the
number of open channels $N_\mathrm{o}$, due to products of the reaction,
is much larger than one in this case.

We will first set the context for complex lifetimes, then re-derive the approach
of Peskin \emph{et al}, with an emphasis on small values of $N_\mathrm{o}$.
The rates will depend on the value of a ``sticking coefficient''
$x$~\cite{croft.bohn.ea:unified}, which is associated with the probability
that a complex is formed upon collision.
Next, the model must be extended to accommodate threshold effects in the
ultracold gases of interest, whereby the observed lifetimes will also be a
function of the molecule-molecule scattering length.
Finally we will apply the model to give insight into the rates observed in
several current experiments.

As a final introductory  note, it may be worth considering the influence of nuclear spins in the complexes.  If the nuclear spins of different hyperfine manifolds are coupled, then these additional degrees of freedom would increase the effective density of states, hence  also increase the RRKM lifetime.  
 The role of nuclear spins has been discussed recently by several auuthors~\cite{mayle.ruzic.ea:statistical,frye.hutson:complexes,Jachymski21_preprint,Hu_NC_13_435_2021,Quemener21_preprint}.
For ultracold exothermic processes such as for KRb molecules, experimental results are consistent with the fact that nuclear spins play a spectator role in the complex \cite{Hu_NC_13_435_2021,Quemener21_preprint}.
As $N_\mathrm{o}$ is large for this type of process, the lifetime of the complex
in Eq.~\eqref{eq:RRKM_lifetime} might be short enough so that spin changing processes do not have time to occur~\cite{Jachymski21_preprint}.
In contrast for ultracold endothermic processes, $N_\mathrm{o} = 1$, lifetimes become longer and the role of nuclear spins might play an important role~\cite{mayle.ruzic.ea:statistical,frye.hutson:complexes,Jachymski21_preprint}.
In any event, we do not consider such an influence explicitly in what follows.

\section{Method and Model}
The physics of the complexes is divided conceptually into two arenas,
the formation of the complex and its subsequent decay,
regarded as independent events.
Central to the theory of complex formation in ultracold molecules is a
statistical approach, which has been adopted from the literature of nuclear
physics~\cite{bohr:neutron,feshbach.weisskopf:schematic,
weisskopf:compound,feshbach.porter.ea:model,friedman.weisskopf:compound,
feshbach:optical,feshbach:unified,feshbach:unified*1} and unimolecular
dissociation~\cite{forst:theory,peskin.reisler.ea:on}.
This approach assumes that many resonances occur in the range of collision energies considered, so that quantities such as the scattering matrix can be averaged over these resonances.
In this way, the averaged scattering matrix becomes sub-unitary, and in the theory it can account for the apparent loss of molecules due to complex formation~\cite{mitchell.richter.ea:random,croft.bohn.ea:unified}.

And here is where the problems begin.
For, the density of states  $\rho$ appears to be not strictly large enough for this
critical assumption of the model to hold.
The mean spacing between energy levels in the complex, $d = 1 / \rho$,
tends to be large on ultracold scales,
ranging from $\sim$~\SI{200}{\micro\kelvin} in LiNa,
to $\sim$~\SI{0.2}{\micro\kelvin} for RbCs~\cite{christianen.karman.ea:quasiclassical}.
Thus in a gas of temperature $T \approx$~\SI{0.5}{\micro\kelvin},
the number of resonances directly relevant would be of order unity at best,
and far below this in general.
Additional resonances could be relevant if they are very broad, but this
remains uncertain at present~\cite{christianen.groenenboom.ea:lossy}.

Nevertheless, molecules in experiments certainly behave \emph{as if} they are
governed by a statistical theory.
Molecular loss is convincingly modeled by theories in which the molecules,
upon approaching a certain relative distance,
vanish with a probability that usually sits somewhere between 0.5 and
1~\cite{gregory.blackmore.ea:molecule-molecule}.
It is assumed that these molecules vanish into complexes,
as there is no place else for them to go.
Indeed, in at least one case, KRb+KRb collisions (where there actually is somewhere else to go), the resulting complexes have
been directly observed by photoionization followed by mass spectrometry~\cite{hu.liu.ea:direct}, as have the products of reaction, which are distributed according to statistical laws~\cite{ Liu21_Nat}.

Moreover, in the case of complexes destroyed by light scattering, this
destruction appears to be adequately described by a model that assumes
photoabsorption by complexes~\cite{christianen.zwierlein.ea:photoinduced}.
Finally, the observed ability of an applied electric field to increase the rate
of loss due to complex formation~\cite{guo.ye.ea:dipolar,
bause.schindewolf.ea:collisions}, is explained using statistical analyses of the complex.
Specifically, the effective density of states rises when the total angular
momentum of the complex is no longer conserved and states of alternative total
angular momentum contribute~\cite{quemener.croft.ea:electric}.

Further evidence of statistical behavior comes from calculations.  For example, classical trajectory calculations certainly seem to indicate that the
atoms in the complex explore phase space ergodically~\cite{croft.bohn:long-lived,
Klos_SR_11_10598_2021}
Moreover, quantum scattering calculations confirm that ultracold
molecular collisions agree with the statistical predictions of quantum
chaos~\cite{croft.makrides.ea:universality,croft.balakrishnan.ea:long-lived,
kendrick.li.ea:non-adiabatic}.

One further argument may be lobbied in favor of the statistical approach.
In quantum statistical mechanics, it is known that an isolated system, evolving according to unitary time evolution, can exhibit apparent thermalization, if the system is chaotic and if only a few degrees of freedom are sampled.
In such a case, thermodynamic variables that should be computed by averaging over \emph{many} quantum states, are actually well-represented by an average over a \emph{single} representative state, since either way, the chaotic states explore large portions of the available phase space.  This is a consequence of the eigenstate thermalization hypothesis~\cite{dAlessio16_PA}.
For the theory of collision complexes, one may run this argument in reverse.
If only approximately a single resonant state is populated in a collision, no matter; we replace averages over this state by an ensemble average over convenient nearby resonant states.

In any event, motivated by these encouraging examples, and in the interest of
developing the statistical analysis as far as possible, we freely incorporate a
statistical approach here and intend to average decay rates over many resonances.

\subsection{Rates in transition state theory}
We therefore begin by positing that a complex has been formed,
and that it has access to considerable phase space,
as  measured in the model by incorporating many resonant states.
The complex has total energy $E$, assumed to lie in the vicinity of the
original entrance channel.
The density of states $\rho(E)$ at this energy defines a characteristic
frequency ${(2 \pi\hbar \rho)}^{-1} = d/(2 \pi \hbar)$, identified by Weisskopf
as the frequency with which a wave packet composed of stationary states of
the complex can reach the complex's periphery and try to
escape~\cite{weisskopf:compound,friedman.weisskopf:compound}.

On each attempt at escape, let the probability of actually escaping to infinity
in open channel $i$ be denoted $p_i$.
If there are $N_\mathrm{o}$ open channels, $i = 1, 2, \dots, N_\mathrm{o}$,
then the escape rate is that given by transition state theory
(TST)~\cite{forst:theory,peskin.reisler.ea:on},
\begin{align}\label{eq:k_TST}
k_\mathrm{TST} = \frac{d}{ 2 \pi \hbar  } \sum_{i=1}^{N_\mathrm{o}} p_i.
\end{align}
In the very simplest approximation, one assumes that the probability is unity in each open channel, $p_i=1$, whereby
\begin{align}\label{eq:k_RRKM}
k_\mathrm{TST} \to \frac{d}{2\pi\hbar} \sum_{i=1}^{N_\mathrm{o}} 1 =
\frac{N_\mathrm{o}d}{2\pi\hbar}  = k_\mathrm{RRKM}.
\end{align}
This gives the RRKM rate, whose reciprocal is of course the RRKM lifetime
 in Eq.\eqref{eq:RRKM_lifetime}.
Thus, in general, $\tau_\mathrm{RRKM}$ represents a lower limit on the lifetime,
at least within transition state theory.

In general however, the escape probabilities are less than unity.
For our purposes, we write
\begin{align}\label{eq:barpespilon}
p_i = {\bar p}_i \epsilon_i.
\end{align}
Here ${\bar p}_i$ is a short-range parameter, governed by the coupling of the
states of the complex to the open channel $i$,
and describing the probability to actually enter the channel;
and $\epsilon_i$ is a factor that accounts for the additional challenge the
outbound wave function faces in  propagating to infinity, incorporating, where
appropriate, the effects of the Wigner threshold laws.
We return to $\epsilon_i$ below.

By detailed balance, the probability ${\bar p}_i$ is the same as the
short-range sticking probability, which in the statistical theory is given by
an energy average of a short-range scattering matrix,
\begin{align}
{\bar p}_i = 1 - | \langle {\bar S}_{ii} \rangle |^2 = \frac{4 x_i}{ (1 + x_i)^2},
\label{eq:pbar_def}
\end{align}
in terms of a dimensionless parameter $x_i$.
This dimensionless parameter given by
\begin{equation}
x_i = \frac{ \pi^2 \nu_i^2 }{ d }.
\label{eq:x_def}
\end{equation}
It is defined in terms of the mean coupling of the resonances to the open channels,
denoted $\nu_i^2$ and the mean spacing~\cite{mitchell.richter.ea:random,croft.bohn.ea:unified}.
More generally, loss of molecules upon colliding is governed by a loss
coefficient that can incorporate losses due to sticking,
direct chemical reaction, or photodissociation.
This coefficient can be extracted from experimental data as a fit parameter in a generic theory of
loss~\cite{idziaszek.julienne:universal,kotochigova:dispersion,
frye.julienne.ea:cold}.

If we consider the case with only one open channel such that there are no
inelastic losses possible and note that the complex cannot decay before one
full period of oscillation we obtain a lower limit on the lifetime
\begin{equation}
  \tau \ge \frac{2\pi\hbar}{d}.
\end{equation}
Noting the lifetime is also given by $\tau = \hbar/\Gamma$ we find for the
ratio of the mean width to the mean spacing
\begin{equation}
  \frac{\Gamma}{d} \le \frac{1}{2\pi},
\end{equation}
known as the Weisskopf estimate.
For a mean decay width $\Gamma = 2 \pi \nu^2 = 2xd/\pi$,
this result corresponds to an upper limit on $x$ of $1/4$,
 beguilingly close to the measured value for RbCs of $x=$0.26(3)~\cite{gregory.frye.ea:sticky}.
Nevertheless, values of $x$ extracted from collision data sometimes exceed this limit, and indeed in the unified approach~\cite{croft.bohn.ea:unified}  values of $x>1$ describe the potentially physically relevant limit of overlapping resonances, the Ericsson fluctuation regime.
Thus it is already clear that the transition state theory may be inadequate here, necessitating the more detailed optical model in the following subsection.

Still, it is useful to see what the transition state theory describes.
In the limit of weak coupling, where resonance widths are much smaller than the
mean resonance spacing, there occur isolated resonances,
so that $x_i \ll  1$ and ${\bar p}_i =  4\pi^2 \nu_i^2/d$
from Eq.~\eqref{eq:pbar_def} and Eq.~\eqref{eq:x_def}.
In this limit the rate in transition state theory becomes
\begin{align}
k_\mathrm{TST} = \frac{d}{2\pi\hbar} \frac{4\pi^2}{d} \sum_{i=1}^{N_\mathrm{o}} \nu_i^2
= \frac{\bar \Gamma}{\hbar} = k_\mathrm{res},
\label{eq:kTST_eq_kRES}
\end{align}
in terms of the mean resonance width
\begin{align}
{\bar \Gamma} = 2 \pi \sum_i \nu_i^2.
\label{eq:mean_Gamma}
\end{align}
This is of course just the decay rate expected for an isolated resonance.

\begin{figure}
\centering
\includegraphics[width=0.99\columnwidth]{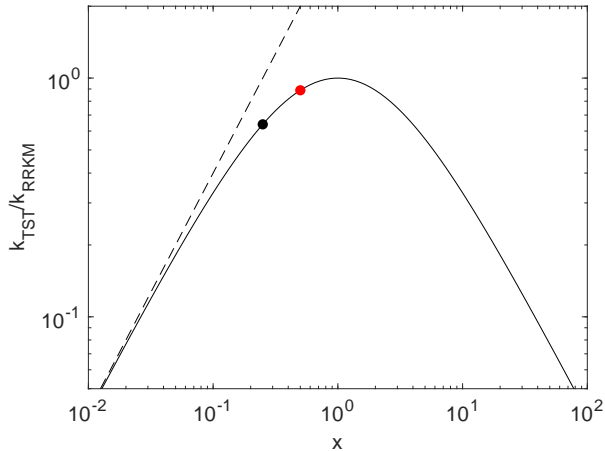}
\caption{
  Solid line: decay rate of the complex $k_\mathrm{TST}$, scaled to the RRKM rate,
  in the simplest transition state theory Eq.~\eqref{eq:k_TST},
  where the sticking coefficient $x$ is assumed to be the same in all open channels.
  Note that $k_\mathrm{TST}$ is equal to the RRKM rate only when $x=1$,
  otherwise is lower.
  The dashed line shows the resonance approximation to the rate $k_\mathrm{res}$,
  Eq.~\eqref{eq:kTST_eq_kRES}, also normalized to $k_\mathrm{RRKM}$, which is expected to be valid only in the
  limit of isolated resonances, $x \ll 1$.
  The dots represent  the results for values of $x$, inferred from data: $x=0.26$  for RbCs (black), and $x=0.5$ for NaRb  (red). }
  \label{fig:TST_rates}
\end{figure}

In what follows, we will allow $x_i$ to be a free parameter whose value can
run between 0 and $\infty$.
Note that the probability ${\bar p}_i$, as defined in Eq.~\eqref{eq:pbar_def}, has
the same value for $1/x_i$ and for $x_i$.
A general overview of rates in the transition state picture is shown in Figure~\ref{fig:TST_rates}.
Here the solid line shows the transition state rate result Eq.~\eqref{eq:k_TST},
assuming for simplicity that the sticking coefficient $x_i=x$ is the same in all channels
with a same probability ${\bar p}$.
Thus the TST rate is $k_\mathrm{TST} = (d / 2\pi \hbar) {\bar p} N_\mathrm{o}$.
This rate is pictured here normalized to the RRKM rate, whereby this ratio is
independent of the number of open channels $N_\mathrm{o}$.
It is therefore clear that $k_\mathrm{TST}$ is in general less than $k_\mathrm{RRKM}$,
becoming equal only in the limit of full absorption $x=1$.
The ratio $k_\mathrm{TST}/k_\mathrm{RRKM}$ becomes small in both the weak-
and strong-coupling limits, that is, at small and large $x$.
Also shown in the figure as the dashed line, is the resonant rate $k_\mathrm{res}$,
also normalized by $k_\mathrm{RRKM}$.
As expected, the resonant rate agrees with $k_\mathrm{TST}$ in the limit of
weak coupling (small $x$), and far exceeds it otherwise.

As a point of reference, the solid black circle denotes the result for $x=0.26$,
close to the measured sticking coefficient for RbCs~\cite{gregory.frye.ea:sticky}.
This simplified transition state theory model would yield in this case,
the ratio $k_\mathrm{TST}/k_\mathrm{RRKM} = 0.64$,
compellingly close to the empirical ratio $0.48 \pm 0.06$ measured in Ref.~\cite{gregory.blackmore.ea:loss}.
This apparent agreement is almost certainly fortuitous,
since for one thing this simple theory does not account for the influence of
the threshold behavior on the decay of the complex.
For another, it flies in the face of other experiments, e.g., NaRb scattering, where the TST rate is comparable to the RRKM rate (red circle in the figure), yet the observed decay rates are
orders of magnitude slower than the RRKM rate, a result not accommodated in
this simple theory.

\subsection{Optical model}
Transition state theory provides approximate rate constants,
based on a key parameter, the attempt frequency $d/(2 \pi \hbar)$ for the
complex to decay, which is a scale, not a specific frequency of particular events.
More concretely, time evolution in quantum mechanics is governed by the relative
energies of the stationary states.
We here develop this picture, based on an optical model
Hamiltonian~\cite{feshbach:optical,feshbach:unified,feshbach:unified*1}.

We begin with the resonant spectrum.
This is described by the Hamiltonian
\begin{align}
H^\mathrm{eff}_{\mu \nu} = E_{\mu} \delta_{\mu \nu} - i \pi \sum_i W_{\mu i}W_{i \nu}.
\label{eq:optical_Ham}
\end{align}
The bare energies $E_{\mu}$ are assumed to be the eigenvalues of a model
Hamiltonian $H^\mathrm{GOE}$ drawn from a Gaussian orthogonal ensemble
with mean spacing $d = 1/ \rho$;
and the coupling constants $W_{\mu i}$ are Gaussian distributed with mean values
\begin{align}
\langle W_{\mu i} W_{\nu j} \rangle = \delta_{\mu \nu} \delta_{ij} \nu_i^2,
\label{eq:mean_width}
\end{align}
that is, $\nu_i^2$ is the variance of the matrix elements.
This value of $\nu_i^2$ defines the dimensionless coupling constant $x_i$ in Eq.~\eqref{eq:x_def}.

This model results in  a short-range $S$-matrix,
\begin{equation}
{\bar S}_{ij}(E) = \delta_{ij} - 2 \pi i \sum_{\mu \nu} W_{i \mu}
\Big( (E-E_{\mu}) \delta_{\mu \nu} + (i/2) \Gamma_{\mu \nu} \Big)^{-1}_{\mu \nu}
W_{\nu j}
\label{eq:short_range_S}
\end{equation}
composed of resonances whose widths can be defined as
\begin{align}
\Gamma_{\mu \nu} = 2 \pi \sum_k W_{\mu k} W_{k \nu},
\end{align}
under the assumption that the coupling constants $W_{\mu k}$ depend weakly on
energy (\cite{mitchell.richter.ea:random}, pp. 2859--60).
The mean resonant width is therefore as given in Eq.~\eqref{eq:mean_Gamma}.

The time evolution of the resonant states is given by the complex eigenvalues
of $H^\mathrm{GOE}$,
\begin{align}
E_{\lambda} - \frac{i}{2} \Gamma_{\lambda} .
\label{eq:optical_eigenvalues}
\end{align}
A state of the collision complex is assumed to be a coherent superposition
with initial probability amplitude $c_{\lambda}$ in resonant state
$\lvert \lambda \rangle$.
In accord with the statistical approximations of the model,
the coefficients $c_\lambda$ are randomly chosen, subject to normalization
$\sum_\lambda |c_\lambda|^2 = 1$~\cite{peskin.reisler.ea:on}.
The time evolution of such a state is
\begin{align}
|\psi(t) \rangle = \sum_{\lambda} c_{\lambda} |\lambda \rangle
\exp\Big[- \frac{ i }{ \hbar }(E_{\lambda} - (i/2) \Gamma_{\lambda}) t \Big].
\end{align}
The time evolution of the diminishing  population is therefore multi-exponential,
\begin{align}
P_\mathrm{SR}(t) =
\langle \psi(t) | \psi (t) \rangle = \sum_{\lambda} |c_{\lambda}|^2 \, \exp(- \Gamma_{\lambda}t/\hbar )
\label{eq:decay_SR}
\end{align}
This is arbitrarily normalized to $P_\mathrm{SR}(0)= \sum_{\lambda} |c_{\lambda}|^2 = 1$, hence represents the time-evolving population of complexes relative to the initial population.  Here we denote this population with the subscript ``SR'' to emphasize that  it describes only the short-range physics of coupling between bound states $\mu$ and continua $i$, without incorporating threshold effects.  
In the optical model, the resonance widths $\Gamma_\lambda$ represent decay
arising from the non-unitarity of the short-range Hamiltonian Eq.~\eqref{eq:optical_Ham}.
In this Hamiltonian, the ability of a given bound state $\mu$ to decay depends
on its coupling $W_{\mu i}$ to all the open channels.  As a result the decay rates in this model depend on the number $N_\mathrm{o}$
of open channels, as we  will see explicitly below.

\subsection{Influence of the threshold}
The optical Hamiltonian we have defined says nothing, however, about
the channels $i$ that the molecules decay into, and their influence must also
be incorporated.
For the sake of simplicity, we account for threshold effects assuming that
there is a single open channel, $N_\mathrm{o}=1$.
This already encompasses some interesting (one might argue, the most
interesting) of the experimental situations.

For a single open channel we drop the channel index $i$.
Within the transition state theory Eq.~\eqref{eq:k_TST} approach,
and in the limit of isolated resonances where Eq.~\eqref{eq:kTST_eq_kRES} holds,
we have
\begin{equation}
k_\mathrm{TST} =  \left(\frac{d}{2\pi\hbar}{\bar p}\right) \epsilon = \frac{\bar\Gamma}{\hbar}\epsilon.
\end{equation}
Here $\bar \Gamma$ represents the mean value of the decay widths for short-range physics,
that is, the mean of the $\Gamma_\lambda$'s arising from the optical model.
Thus for the purpose of time evolution of the decaying complex, the optical model
rates $\Gamma_\lambda$ are modified by a threshold correction factor $\epsilon(E)$,
 as introduced in Eq.~\eqref{eq:barpespilon},
which extends the lifetimes of the complexes due to quantum reflection back
to short range off the long range potential.
We refer to the decay that includes this long-range physics as
\begin{align}
P_\mathrm{LR}( t) = \sum_{\lambda} |c_{\lambda}|^2 \exp(-\Gamma_\lambda\epsilon t/\hbar).
\label{eq:NLR}
\end{align}
As such $\epsilon$ can be seen to simply be a re-normalization of the mean
coupling strength defined in Eq.~\eqref{eq:mean_width} to account for threshold
effects,  which $P_\mathrm{SR}$ does not take into account.

The form of $\epsilon(E)$ is quantified in terms of standard quantum defect
theory parameters $C^{-2}$ and $\tan\lambda$ in the open channel,
both energy dependent.  The modification coefficient is given
by~\cite{croft.bohn.ea:unified}
\begin{align}
\epsilon(E)  = \frac{C^{-2}(1+x)^2}{(1+x C^{-2})^2 + x^2 (\tan \lambda)^2}.
\label{eq:epsilon_def}
\end{align}
For $s$-wave scattering, the MQDT parameters are parametrized as follows~\cite{idziaszek.julienne:universal}
\begin{align}
C^{-2} &= k{\bar a} \left[ 1 + (s-1)^2 \right] \\
\tan \lambda &= 1-s,
\end{align}
where $k  =  \sqrt{ 2 \mu E / \hbar^2}$ is the wave number of the decaying molecule
pair and ${\bar a}$ is the usual Gribakin-Flambaum mean scattering length.
The parameter $s = a/{\bar a}$ is the scattering length in units of ${\bar a}$.

The set of possible collision energies is assumed to coincide with the
original distribution of collision energies $f(E)$ from which the complexes
were originally formed.
The decay of the relative population, including threshold effects, is therefore
given by
\begin{multline}
P_\mathrm{therm}( t) =  \int dE f(E) P_\mathrm{LR}(t)    \\
= \int dE f(E) \sum_{\lambda} |c_{\lambda}|^2
  \exp{(-\Gamma_{\lambda} \epsilon(E) t/ \hbar )}.
\label{eq:Nthresh}
\end{multline}
In this notation $N_\mathrm{o}$ is omitted, since it is assumed to be unity.
In the cases considered below, we assume the distribution of collision energies
is given by the Maxwell-Boltzmann distribution,
\begin{align}
f(E) dE = \frac{2\beta^{3/2}}{\pi^{1/2}} E^{1/2} e^{-\beta E} dE,
\end{align}
where $\beta = 1/k_B T$ for a temperature $T$.

\section{Short-range suppression of decay}
In this section we will first focus on the short-range relative population
function Eq.~\ref{eq:decay_SR}, and disregard threshold effects.
This will allow us to examine the effects of the distribution of the widths
within the optical model on the overall decay, and in particular to emphasize the significance of having only a single open channel.

\subsection{The chi-square distribution}
As a preliminary, we note that the distribution of widths that results from the optical model is a known quantity, at least in the limit of small $x$.
For a given  number $N_\mathrm{o}$ of open channels,
the distribution of widths $\Gamma$ is given by the chi-squared
distribution~\cite{porter.thomas:fluctuations,weidenmuller.mitchell:random,mitchell.richter.ea:random}
\begin{equation}\label{eqn:chi2}
  P_{\chi^2}(\Gamma) = \left(\frac{N_\mathrm{o}}{2\bar\Gamma}\right)^\frac{N_\mathrm{o}}{2}\frac{\Gamma^{N_\mathrm{o}/2 - 1}}{ [\Gamma\left(N_\mathrm{o}/2\right)]}\exp\left({-\frac{N_\mathrm{o}\Gamma}{2\bar\Gamma}}\right).
\end{equation}
(Here it should be emphasized that the factor $\Gamma(N_\mathrm{o}/2)$  in the square brackets of the denominator of the second term refers to the ordinary gamma function, evaluated at $N_o/2$, and does not pertain to the width $\Gamma$).
 The mean of this distribution is given by ${\bar \Gamma} = 2 \pi \nu^2 N_o$, thus the {\it mean} decay rate would simply scale linearly in the number of open channels, just as in the RRKM theory.

\begin{figure}
\centering
\includegraphics[width=0.99\columnwidth]{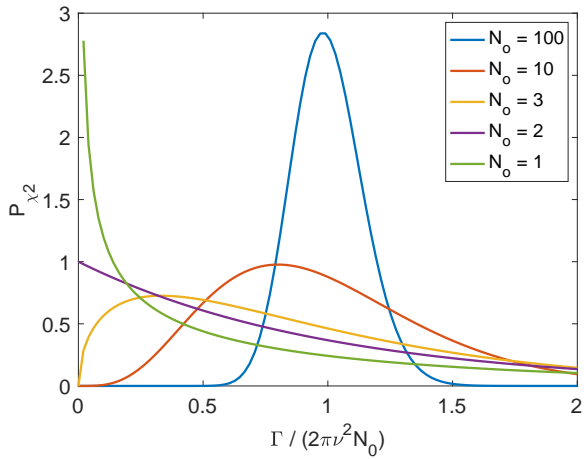}
  \caption{The chi-squared distribution for the widths, Eq~\eqref{eqn:chi2}.
  It is clearly seen that when the number of open channels becomes small, each
  distribution has its own mercurial character.
  Whereas when the number of open channels becomes large the distribution tends
  to a normal distribution centered at the mean width ${\bar \Gamma} = 2 \pi \nu^2 N_\mathrm{o}$.}
  \label{fig:chi2}
\end{figure}

However, the detailed distribution of widths shows a tremendous qualitative difference between large and small numbers of open channels, as illustrated in Figure~\ref{fig:chi2}.
This figure plots $P_{\chi^2}$ as a function of the width normalized to the mean width ${\bar \Gamma}$, for various values of $N_\mathrm{o}$.  For many open channels, $N_\mathrm{o}=100$, the distribution of widths is peaked around the mean, whereby the mean width is a reasonable stand-in for any of the widths.  In the other limit, $N_\mathrm{o}=1$, the distribution is instead peaked strongly toward widths far smaller than the mean width, and indeed the distribution diverges as $\Gamma^{-1/2}$ for small $\Gamma$.

This difference between large and small $N_o$ in deciding the width distribution is critical to the theory of the decay, and forms the key observation of the paper.
We may think of it somewhat intuitively by considering the perturbative limit.
First, consider the case of a single open channel $i=1$.
The coupling matrix elements $W_{\mu i}$ are drawn from a normal distribution with zero mean and variance $\nu^2$.
Denoting the random variable as $w = W_{\mu i}$, the distribution of coupling matrix elements is  normal by definition,
\begin{align}
P_\mathrm{couplings}(w) \propto \exp \left( - \frac{ w^2 }{  2 \nu^2 } \right),
\end{align}
where we do not require the normalization for our present purposes.

Each resonance $\mu$ decays with a width given, in this perturbative limit, by a distribution of values $\Gamma = 2 \pi w^2$.
The distribution of widths therefore is given by the distribution of the squares of the random variable $w$.
By the usual change of variable rules, this gives the distribution
\begin{align}
P_\mathrm{widths}( \Gamma ) & = P_\mathrm{couplings}(w) \left( \frac{ d \Gamma }{ dw } \right)^{-1} \nonumber \\
& \propto \exp \left( - \frac{ w^2 }{ 2 \nu^2 } \right) \frac{ 1 }{ 4 \pi w }  \nonumber \\
& \propto \exp \left( - \frac{ \Gamma }{ 4 \pi \nu^2 } \right) \frac{ 1 }{ \sqrt{ \Gamma} },
\end{align}
which is of course the functional form when $N_\mathrm{o}=1$ in Eq.~\eqref{eqn:chi2}.  The point is that the coupling matrix elements are clustered around zero.  Since the values of $w/\nu$ are predominantly less than unity, then their squares must be clustered even closer around zero.  This is expressed in the $1 / \sqrt{\Gamma}$ divergence in the width distribution.  From this perspective, it is perfectly natural that the single channel case is dominated by short widths and long lifetimes.

In spite of this heavy preference for small values of $\Gamma$, nevertheless the mean of the distribution for $N_\mathrm{o}=1$ is still $2 \pi \nu^2$.  Thus for many open channels, where  in the perturbative limit the total decay rate is the sum
\begin{align}
\Gamma = \sum_{i=1}^{N_o}  \Gamma_i = \sum_{i=1}^{N_o} 2 \pi  w_i^2,
\end{align}
the average value of each term is $2 \pi  \nu^2$ and thus the sum must have mean value ${\bar \Gamma} = 2 \pi \nu^2N_\mathrm{o}$.   Even though any given $w_i^2$ is likely to be near zero, the more such terms there are, the less likely it is that the whole sum is near zero.  In particular, the probability of finding any given $\Gamma_i$ within $\delta \Gamma$ of zero is proportional to
\begin{align}
\int_0^{\delta \Gamma} d\Gamma  \exp( - \Gamma / 2 \pi \nu^2 ) \Gamma^{-1/2}
\approx ( \delta \Gamma )^{1/2},
\end{align}
whereby the probability that all the rates are smaller than $\delta \Gamma$ scales as $(\delta \Gamma)^{N_o/2}$, vanishingly small in the limit of large $N_o$.  Note this is the same scaling as the chi-squared distribution, integrated to a small $\delta \Gamma$.   In the opposite limit of many open channels we note that the mean of each width distribution $\Gamma_i$ is the  $2 \pi \nu^2$.  The mean of the total width $\sum_i \Gamma_i$ is then ${\bar \Gamma}$, and the distribution is peaked around this value.

The preponderance of small $\Gamma$'s in the $N_\mathrm{o}=1$ limit has serious consequences for the time dependent decay.
To see this, we evaluate the sum in Eq.~\eqref{eq:decay_SR}.
To make this explicit, we assume the population $|c_{\lambda}|$ of each resonance is proportional to the probability $P_{\chi2}(\Gamma_{\lambda})$ that the corresponding width $\Gamma_{\lambda}$ occurs.  In this case, Eq.~\eqref{eq:decay_SR} can be re-interpreted as the integral
\begin{align}
  P_\mathrm{SR}(t) &\approx \int_0^\infty \diff\Gamma P_{\chi2}(\Gamma)\exp(-\Gamma t/\hbar) \nonumber  \\
                 &= {\left(1 + \frac{2\bar\Gamma t}{\hbar N_\mathrm{o}}\right)}^{-\frac{N_\mathrm{o}}{2}}.
\end{align}

In the limit $N_\mathrm{o} \to \infty$, this expression gives the exponential decay law at the mean rate ${\bar \Gamma}$,
\begin{equation}\label{eqn:exp}
 P_\mathrm{SR}(t) \rightarrow \ee^{-\bar\Gamma t/\hbar},
\end{equation}
which corresponds to the case of reactive molecules such as KRb which should
therefore exhibit an exponential decay.

However, in the single channel case the expression becomes
\begin{equation}\label{eqn:slow}
  P_\mathrm{SR}(t) = \frac{1}{\sqrt{2 (\bar\Gamma /\hbar) t  + 1}},
\end{equation}
representing an algebraic, rather than exponential, decay.
It agrees with the exponential decay at short times, but at time scales $\approx \hbar /{\bar \Gamma}$ the decay begins to slow down.

Within the RRKM theory, it is assumed that the mean width is representative of the distribution of widths.
However we can see in Figure~\ref{fig:chi2} this is not always the case.  In particular, when there is only
one open channel, the most likely width is in fact zero.
Therefore, any approaches that make this assumption will be unable to capture
the rich physics of the one open channel case.  In particular, this effect was ignored in early work that attempted to describe complex lifetimes using the simpler RRKM theory \cite{mayle.ruzic.ea:statistical,mayle.quemener.ea:scattering}.

\subsection{Numerical examples}

To illustrate this effect we posit a generic numerical model with $N_c=1000$
states in the closed channel space, and a mean energy spacing $d$
that defines the unit of energy.
Doing so also identifies a unit of time $\hbar/d = 1/d$, where we arbitrarily
ignore the value of $\hbar$.
This schema is justified, as we intend to compare ratios of decay rates rather
than absolute rates.
Further, we define a sticking coefficient $x$ that defines the mean coupling
strength $\nu^2 = (d/\pi^2)x$ according to Eq.~\eqref{eq:x_def}.
Finally---and crucially---we define the number of open channels $N_\mathrm{o}$.
We then construct the optical model Hamiltonian Eq.~\eqref{eq:optical_Ham} and find
its eigenvalues Eq.~\eqref{eq:optical_eigenvalues}.
This process defines an ensemble of resonance widths $\Gamma_{\lambda}$,
in terms of which the decay of the population is given by
$P_\mathrm{SR}(t)$  in Eq.~\eqref{eq:decay_SR}.

\begin{figure}
\centering
\includegraphics[width=1.05\columnwidth]{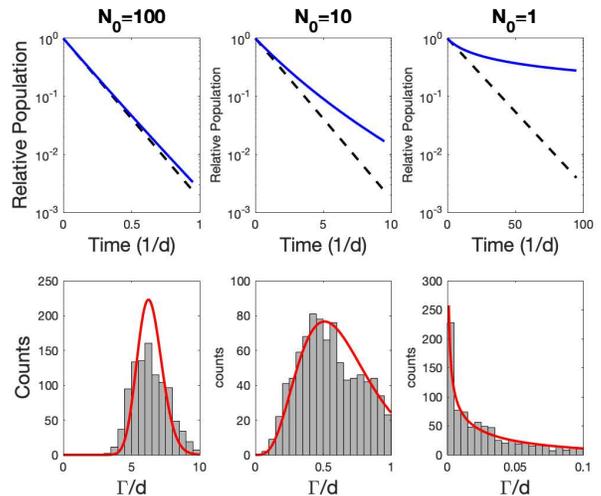}
\caption{Characteristic decay of the initial population (top row) and
  distribution of resonance width (bottom row) for the model described in the
  text, with $x=0.1$ and the three numbers of open channels $N_\mathrm{o}$ given. In each upper panel, the dashed lines give the  decay $\exp(-{\bar \Gamma}t)$ at the mean decay rate, while the blue solid lines give the decay according to the optical model.  In the numerical width distributions shown in the bottom row, the red line indicated the chi-square distribution Eq.~\eqref{eqn:chi2}.}
  \label{fig:optical_model_decay}
\end{figure}

The model is perhaps most easily interpreted in the weak-coupling limit,
$x \ll 1$, where the transition state theory rate is predicted to nearly
coincide with decay at the mean rate ${\bar \Gamma}$, as described above.
This decay would be at the approximate rate
\begin{equation}
\bar \Gamma = 2\pi N_\mathrm{o} \nu^2 = \frac{2}{\pi}  d N_\mathrm{o} x,
\end{equation}
which provides a reference with which to compare decay including the
suppression due to the distribution of widths.
We therefore first consider a model with $x=0.1$, where the mean decay rate,
in units of mean spacing, would be $\bar \Gamma/d = (2/\pi) 0.1 N_\mathrm{o} \approx 0.064 N_\mathrm{o}$.

Examples of this decay, for $N_\mathrm{o}=100$, $10$, and $1$ open channels,
are shown in the three columns of Figure~\ref{fig:optical_model_decay}.
The decay for $N_\mathrm{o}=100$ open channels is shown in the top left panel.
Here, the optical model decay (solid blue line) is in fairly good agreement with the
exponential decay at the mean rate, $\exp(-{\bar \Gamma}t)$ (dashed line).
This is because all the rates are narrowly distributed
around the mean rate ${\bar \Gamma}/d = 6.4$, as seen in the lower left panel.

For a smaller number of open channels, $N_o=10$ (middle column), the overall decay is ten times slower than for $N_\mathrm{o}=100$, as expected on general grounds, e.g., the RRKM theory.
In this case, the agreement of the optical model decay with the mean decay is good at
relatively short times, but not so good later (upper middle).
Here the widths have a broader distribution around the mean ${\bar \Gamma}/d = 0.64$
(lower middle) and in particular this distribution samples low values of $\Gamma$,
leading to a slow long-time decay.

Finally, in the extreme case of a single open channel (right panel), the mean decay is yet another order of magnitude slower.
It is clear that the decay, while starting similarly to the exponential decay, nevertheless quickly deviates from it,
and is no longer well represented by an exponential decay on any relevant timescale.
In this case the blue curve showing decay in the optical model is well-represented by the analytical approximation Eq.~\eqref{eqn:slow}.
Again, this arises from a distribution of widths that shows high incidence of resonances with minuscule width as compared to the mean width (lower right panel).

\subsection{Rates versus coupling strengh}

This striking additional longevity of the complex for small $N_\mathrm{o}$ is
illustrated in Figure~\ref{fig:optical_model_decay} in the weak-coupling limit, $x=0.1$,
but it is generic for all values of $x$.
To show this, we would like to plot a decay rate versus $x$.
This is somewhat problematic, since, by its very nature the decay is
non-exponential, hence not easily characterized by a number $k_\mathrm{eff}$ in an expression $\exp(-k_\mathrm{eff}t)$.
To circumvent this difficulty, we somewhat arbitrarily fit the 
population $P_\mathrm{SR}(t)$ to a single decay rate over the time interval
over which $P_\mathrm{SR}$ drops to $1/e^2$ of its initial value.
The linear fit
\begin{align}
  \ln P_\mathrm{SR}(t) = - k_\mathrm{eff} t + b
\label{eq:keff_def}
\end{align}
over this time interval then identifies an approximate, effective decay rate
$k_\mathrm{eff}$ that at least sets the scale of the observed decay.

\begin{figure}
\centering
\includegraphics[width=0.99\columnwidth]{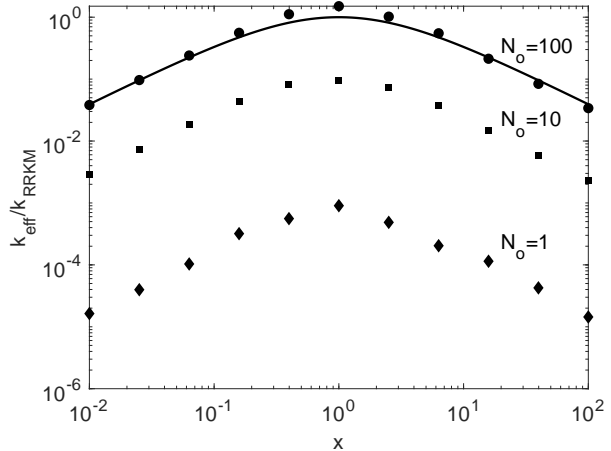}
\caption{Effective decay rates $k_\mathrm{eff}$as defined by
Eq.~\eqref{eq:keff_def}, normalized by the RRKM rate,
  versus sticking coefficient $x$, for various numbers of open channels
  $N_\mathrm{o}=100$ (circles), 10 (squares), 1 (diamonds).
  Solid line: transition state theory rate $k_\mathrm{TST}$.}
  \label{fig:comparative_rates}
\end{figure}

Defined in this way, the effective decay rates are plotted versus $x$ in
Figure~\ref{fig:comparative_rates}, for the three different numbers of open
channels $N_\mathrm{o} = 100$, $10$, $1$ used previously.
Also reproduced for reference is the transition state theory rate $k_\mathrm{TST}$
from Figure~\ref{fig:TST_rates}.
It is immediately clear that for a large number of open channels
$N_\mathrm{o}=100$, the rate derived from the optical model nearly reproduces
 that from transition state theory,
thus validating the use of this theory in cases such as KRb+KRb scattering
where the number of open channels is indeed large.

The decay rate may be expected to decrease linearly in the number of open
channels, as implied by the RRKM formula  in Eq.~\eqref{eq:k_RRKM}.
Recall that in Figure~\ref{fig:comparative_rates} this dependence is already
accounted for, as $k_\mathrm{eff}$ is normalized to $k_\mathrm{RRKM}$.
Thus the figure illustrates an extremely rapid \emph{additional} decline in
$k_\mathrm{eff}$ as $N_\mathrm{o}$ decreases.
Strikingly, this suppression below $k_\mathrm{TST}$ appears to be a similar
factor for all values of $x$.
For $x$ in a range something like $0.1 < x < 10$, and for a single open channel,
a suppression of decay rates by 3--4 orders of magnitude is not unreasonable.
This is intriguing news for observations of collision complexes in NaK+NaK, NaRb+NaRb,
or Rb+KRb collisions, where such orders of magnitude increase in apparent lifetimes
have been observed.

Nevertheless, a direct comparison is difficult to make, as the rates in Figure (\ref{fig:comparative_rates}) are derived using an effective fit (\ref{eq:keff_def}) to a non-exponential curve of the form (\ref{eqn:slow}), rather that to an exponential fit.  We will see the difference in these time-dependent decays below.

\section{Application to experimental systems}

To investigate how well the model works for realistic molecules,
we will incorporate the actual densities of states as estimated in
Ref.~\cite{christianen.karman.ea:quasiclassical} for two representative
species of current experimental interest.
We will analyze corrections to the decay rates according to the correction
factor $\epsilon$ in Eq.\eqref{eq:epsilon_def} as well as due to the distribution
of widths.
Finally, we will average the decay rates over the thermal population of molecules,
to present the relative decays $P_\mathrm{them}(t)$ as given in Eq.~\eqref{eq:Nthresh}.
We will focus on the singular case with only one open channel.

\subsection{RbCs + RbCs}

\begin{figure}[tbh]
\centering
\includegraphics[width=1.2\columnwidth]{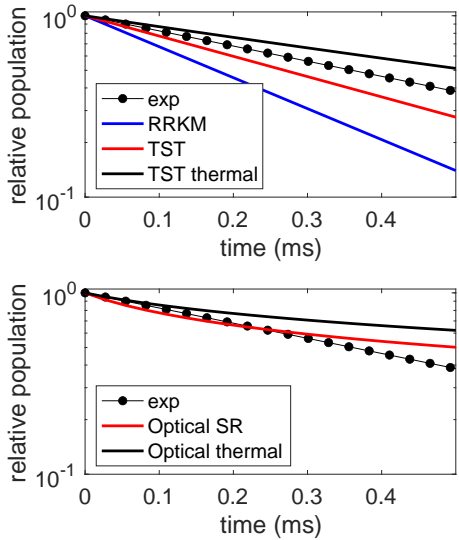}
\caption{Model decays for the (RbCs)$_2^*$ complex at various levels of approximation.
  In each panel the dotted line represents the empirical decay, given by
  $P = \exp(-t/\tau_\mathrm{expt})$, with $\tau_\mathrm{expt} =$ \SI{0.52}{\milli\second}. In the upper panel are various approximations to decays in transition state theory, as described in the text.
 In the lower panel, the optical model decays are considered, both in a purely short-range version (red) and in the complete model that includes threshold effect and thermal averaging (black).
  These results assume $x=0.26$ and $s=1.2$~\cite{gregory.frye.ea:sticky}, and a gas temperature of
  $T = $ \SI{2}{\micro\kelvin} }
  \label{fig:RbCs_decay}
\end{figure}

We begin with the decay of complexes formed in ultracold collisions of RbCs molecules,
as reported in Ref.~\cite{gregory.blackmore.ea:loss},
where the empirical lifetime was determined to be $\tau_\mathrm{expt} =$ \SI{0.52}{\milli\second}.
As noted above, this is already comparable to the RRKM decay rate.
One hopes, therefore, that the influence of the threshold does not wreck this happy situation.
Figure~\ref{fig:RbCs_decay} illustrates the decay of population in various scenarios.
As a stand-in for the experimental data, the dash-dot line is simply the decay
$\exp(-t / \tau_\mathrm{expt})$, plotted on a semi-log scale in each panel.
We will refer to this reference curve as the empirical decay.
Before we proceed we note that the experimental measurement of the lifetime
for RbCs assumes that the decay is exponential which as we have seen is
potentially not true, however does characterize the empirically observed decay in much
the same way as the effective decay we defined earlier does.

The model requires the values of the parameters $x$ (denoting channel coupling)
and $s$ (denoting the scattering length).
Luckily, these have been measured in the original experiment that observed sticking
of RbCs molecules~\cite{gregory.frye.ea:sticky}.
In this experiment, $x$ had the non-universal value $x = 0.26$,
while the scattering length was presented in terms of the short range phase
shift $\delta^s = 0.56 \pi$ as $s = a/{\bar a} = 1 + \cot(\delta^s = \pi/8) = 1.21$.
To fully compute the threshold factor $\epsilon$, we need the  van der Waals
coefficient $C_6 = $ \SI{1.41e5}{\hartree\bohr^6}~\cite{gregory.frye.ea:sticky}.
For thermal averaging, we assume a temperature $T =$ \SI{2}{\micro\kelvin}
typical of the experiment.

In the upper panel of Figure~\ref{fig:RbCs_decay}, decays are computed at various levels of transition state theory.  The blue curve is the very simplest, RRKM version, which decays with lifetime  $\tau_\mathrm{RRKM} = $  \SI{0.25}{\milli\second}, faster than the empirical decay.  This discrepancy is repaired somewhat in the red curve,
which shows the rate $k_\mathrm{TST}$ derived from transition state theory.
This has the effect of reducing the RRKM rate by a factor ${\bar p} = 4x/(1+x)^2 = 0.66$.
Thus $k_\mathrm{TST} = 0.66 \times k_\mathrm{RRKM}$ and the lifetime predicted in
this model is $\tau_\mathrm{TST} = $ \SI{0.39}{\milli\second}, closer to the
measured $\tau_\mathrm{expt} = $  \SI{0.52}{\milli\second} than is $\tau_\mathrm{RRKM}$.
Finally, the black curve incorporates also the influence of the threshold factor $\epsilon$ of Eq.~\eqref{eq:epsilon_def} (black curve).
This result is also thermally averaged at the presumed temperature  \SI{2}{\micro\kelvin}.  The addition of these threshold effects slows the decay further, in fact making decay in this model slower than the experimental data would suggest.

Next, we consider what the decay looks like in the  more sophisticated optical model, depicted in the lower panel of Figure~\ref{fig:RbCs_decay}.  Results are shown for the short-range version ($P_\mathrm{SR}$, red) and including  the influence of the threshold effects and thermal averaging ($P_\mathrm{therm}$, black).  The comparison of the optical model results to the presumed empirical decay (dash-dot)
is problematic, as the latter is assumed exponential while the former is certainly not.  We find that the short-range optical model follows the empirical model fairly well up to approximately 0.3  ms.  However the final model  is in agreement only at the very shortest times, and decays far
more slowly than the simple empirical exponential after about \SI{0.1}{\milli\second}.
This result is to be expected, since, by fighting its way past the long-range potential,
the outbound wave function's escape probability is reduced further,
beyond the influence of short-range physics.


\subsection{NaRb + NaRb}

The situation is somewhat different for other molecular systems whose complex lifetime has been measured recently, namely,  NaRb and NaK molecules~\cite{gersema.voges.ea:probing}.  Here, the measured lifetimes  exceed the RRKM lifetimes by several orders of magnitude.
It is impossible to compare to these observations directly, since in these
experiments, only lower limits to the lifetimes were measured.

\begin{figure}[tb]
\centering
\includegraphics[width=1.2\columnwidth]{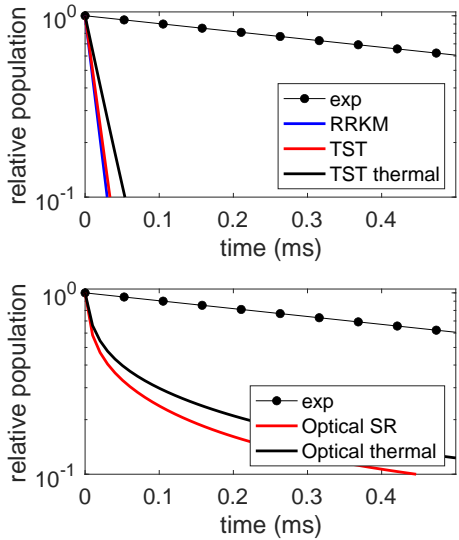}
\caption{Model decays for the (NaRb)$_2^*$ complex, plotted as in Figure~\ref{fig:RbCs_decay}.  
  Dot-dashed line represents an empirical decay, modeled simply as $P = \exp(-t/\tau_\mathrm{expt})$,
  with $\tau_\mathrm{expt} = $ \SI{1}{\milli\second}.  The upper panel shows rates in transition state theory, while the lower panel shos rates from the optical model, using
 $x = 0.5$  and $s=5$.}\label{fig:NaRb_decay}
\end{figure}

For concreteness, we focus on the case of NaRb.
This is because, from the original sticking data in Ref.~\cite{ye.guo.ea:collisions},
the best-fit sticking and phase parameters $x=0.5$ and $s=5$ were inferred~\cite{bai.li.ea:model}.
We will use these nominal values.
The decay was found to have a lifetime in excess of \SI{1}{\milli\second}~\cite{gersema.voges.ea:probing},
so for convenience we compare theory to an exponential decay with lifetime $\tau_\mathrm{expt} =$ \SI{1}{\milli\second}.
Moreover we assume a temperature of \SI{500}{\nano\kelvin}, as in the experiment.
Using these parameters,  we plot  decay profiles in Figure~\ref{fig:NaRb_decay}.
As before, the dot-dashed line is a stand-in for the experimental decay.  We plot here the same set of decay rates, in the same notation, as in Fig.~\ref{fig:RbCs_decay}.  Namely, the upper panel shows the RRKM result (blue), along with plain ($P_\mathrm{SR}$, red) and thermally averaged ($P_\mathrm{therm}$, black) results from transition state theory.  The lower panel shows the optical model result, both in the short-range version (red), and fully thermally averaged (black).  

The point here is that transition state theories predict decay with far shorter lifetimes than experimental decays.  The optical model produces somewhat longer lifetimes, but still not nearly long enough to compare favorably with the data.  

\section{Varying $x$}

As practiced here, the optical model of decay is therefore capable of both overestimating and underestimating the empirical lifetimes.  The calculations of the previous section are performed under a constraint, namely, that the model parameters $x$ and $s$ used in the decay model are the same as those inferred from the formation of the complex.

It is not completely clear that this should be the case.
The parameter $x = \pi^2 \nu^2/d$ has nominally the same origin in the two cases, given by the ratio of mean coupling between closed and open channels, to the mean spacing.
Yet, the role of this parameter in taking averages in the statistical theories is different.
The probability of initial complex formation, ${\bar p} = 4x/(1+x)^2$, relies directly on mean values of the coupling matrix elements, via Eq.~\eqref{eq:mean_width}.
By contrast, for use in the decay, these same couplings must be converted to decay rates $\Gamma_{\lambda}$, and an average is taken over time decays as in Eq.~\eqref{eq:Nthresh}, which, as we have seen, emphasizes values of $\Gamma_{\lambda}$ far less than the mean.  Suppose then that it is not the case that the coupling matrix elements $W$ are Gaussian distributed, but show some additional structure.  Then the two theories, of complex formation and decay, would distort the truth in distinctive ways.  This is a question that will be resolved only upon further understanding of the detailed Hamiltonian of the complex.  In the meantime, we skirt this issue by leaving $x$ as a free parameter.  

The link between parameters and observables is further muddied because, as noted above, it is unclear whether sufficiently many resonances can be accessed for the mean values to be meaningful.  That is, in the limit where averages are taken over a small number of resonances, results may be more susceptible to fluctuations about the mean, and higher-order statistics may be in order.

Finally, the difference between formation and decay of complexes may rely critically on the role of the spin degrees of freedom.  That nuclear spins may play an essential role has been promoted by various authors \cite{mayle.ruzic.ea:statistical,frye.hutson:complexes,Jachymski21_preprint}.  In the present context for ultracold endothermic systems where $N_\mathrm{o} = 1$, consider the possibility
that nuclear spins are initially spectators to the four-atom complex, becoming engaged only after a certain time as
multiple curve crossings occur connecting different spin states.  In this case,  the density of states for decay
would be boosted by an additional factor corresponding to the nuclear spins involved, thus the effective $x$
could be larger for decay than for formation.
Note that for ultracold exothermic systems like KRb,
recent experimental data are consistent
with nuclear spins being spectators \cite{Hu_NC_13_435_2021,Quemener21_preprint}.
But because $N_\mathrm{o} \gg 1$ in these experiments, the lifetimes of the complex are expected to be too short
so that changes of nuclear spins in the complex have  no time to occur, in contrast with
the ultracold endothermic systems.
%

Based on these considerations, it therefore seems reasonable to consider $x$ as a fit parameter for describing the decay of complexes.
We do not, however, attempt to fit any data.
For one thing, no such data exist for the NaRb and NaK experiments, which only set a lower limit for the lifetime.
For another, decays previously measured in RbCs and KRb are extracted from fits to loss data where the optical dipole trap light is chopped, and are modeled by what are assumed to be exponential decays.  Fits using our model would instead assume the algebraic form of decay  $(2({\bar \Gamma} / \hbar) t+1)^{-1/2}$, which would necessitate a fresh look at the data.  While we encourage such fits to be performed, nevertheless we will here content ourselves with finding the appropriate time scales as derived from exponential decays.

\begin{figure}
\centering
\includegraphics[width=3.5in]{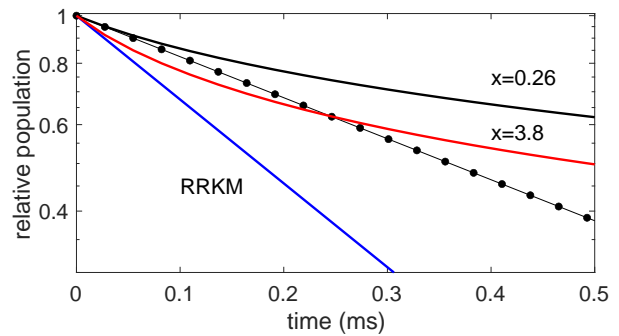}
\caption{Model decays for the (RbCs)$_2^*$ complex.
  Dot-dashed line represents an empirical decay, modeled simply as $P = \exp(-t/\tau_\mathrm{expt})$,
  with $\tau_\mathrm{expt} = $ \SI{0.52}{\milli\second}, as before.
  Upper solid curve is the decay with $x=0.26$, reproduced from Figure~\ref{fig:RbCs_decay}.
  Lower solid line is decay assuming $x = 1/0.26 = 3.8$.}\label{fig:RbCs_two_xs}
\end{figure}

To this end, we revisit the RbCs and NaRb results of the previous section, but freely varying the value of $x$.   As an illustration, we contemplate first RbCs.  In this case the value $x=0.26$ was extracted from complex formation data, assuming that $x \le 1$.  But in the statistical picture of complex formation, values of $x>1$ are also possible, in the limit of overlapping resonances.  Moreover, the short-range probability ${\bar p}$ has a symmetry, it is the same for $1/x$ as for $x$.  Such a symmetry does not apply to the threshold correction factor $\epsilon$.  Thus, merely as an example, we can contemplate a decay model with $x=1/0.26=3.8$, that has the same short-range coupling for formation and decay, but varies only at long range.

Therefore, the observed decay can be different for the two values of $x$, even though they connote the same short-range sticking probability.  This is illustrated in Figure~\ref{fig:RbCs_two_xs}.  For simplicity, this calculation retains the same value of $s=1.2$ for both curves.  Also shown is the RRKM decay for reference.  This example serves merely to illustrate that in this case, decay data may be better explained (the agreement is admittedly completely qualitative!) with larger values of $x$, that is, in the limit of overlapping resonances.  Such a circumstance would also go some way to explaining the observed statistical behavior
despite the seemingly too low density of states, as has been suggested by
Christianen et al.~\cite{christianen.groenenboom.ea:lossy}.

For NaRb, the situation is more drastic.  As we have seen in Figure~\ref{fig:NaRb_decay}, the optical model theory using the values $x=0.5$, $s=5$ inferred from complex formation, underestimates the empirical decay by an order of magnitude or more.  But now, we allow ourselves the possibility of varying $x$ to find better qualitative agreement.

\begin{figure}
\centering
\includegraphics[width=3.5in]{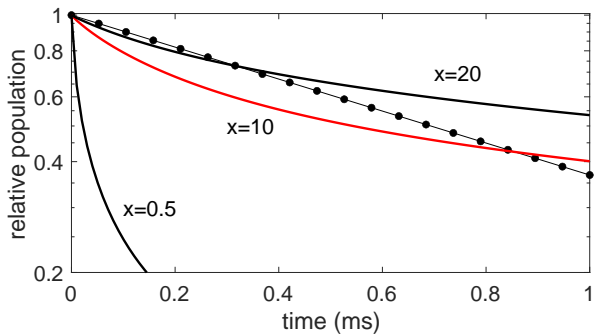}
\caption{Model decays for the (NaRb)$_2^*$ complex.
  Dot-dashed line represents the empirical decay, $P = \exp(-t/\tau_\mathrm{expt})$,
  with $\tau_\mathrm{expt} = $ \SI{1}{\milli\second}.
  The solid curves employ the values of $x$ shown.
  The reduced scattering length is fixed at $s=5$ throughout.}\label{fig:NaRb_three_xs}
\end{figure}

Figure~\ref{fig:NaRb_three_xs} explores just this possibility.
Here we vary the values of $x$ while, for simplicity, fixing the value of $s=5$ already used.
Shown, as always, is the dot-dashed line describing the empirical decay.
Also reproduced is the curve for $x=0.5$ from Figure~\ref{fig:NaRb_decay}.
Non-exponential decay curves in the full model are also shown for the much
larger values $x=10$, $20$.
These decay curves, albeit non-exponential, more realistically describe the
reduction in population over this time scale.



It would perhaps be unwise to pursue this line of thinking more quantitatively,
not least because there is not a measured decay curve to try to fit.
Nevertheless, it is significant that decay curves can be produced in the
present theory that are consistent with very long lifetimes.
In the case of NaRb, this requires a large value of $x$, which, additionally,
strongly suppresses the threshold correction factor $\epsilon$.

\section{Conclusions}

It will be remembered that our overall goal here is to explore what the statistical theory might have to say about complexes forming and subsequently decaying in ultracold alkali dimer collisions. This we do in spite of doubts about the very applicability of such a theory, given the paucity of states within the $\sim k_BT$ energy range available. Nevertheless, the existence of the complexes, their essential randomness, and their utility in thinking about the problem apear no longer in doubt, as we have outlined in Sec. II.

In this article we have added to the qualitative interpretive power of statistical models, namely, to account for  the apparently anomalously long lifetimes of the complex, as compared to the simple standard reference afforded by the RRKM decay rate.
The RRKM rate sets a characteristic scale for the decay rate, which is an upper bound on the true decay rate.
We have found that the decay rate is reduced due to two additional factors.
The first, as might be expected, arises from threshold behavior of the scattering wave function at low energies---the ``quantum reflection on the way out,'' if you will.

The second effect is fundamental to the statistical nature of the complexes, which notes that for small numbers of open channels $N_\mathrm{o}$, the rate is not simply proportional to $N_\mathrm{o}$, as in the RRKM theory, but becomes far slower than this scaling, dramatically so for $N_\mathrm{o}=1$.
This slowing down is intimately tied to a preponderance of unusually long-lived resonant states in the optical model, as codified in the chi-squared distribution, and that is characteristic of a chaotic system. As a consequence, the population of complex states versus time is expected to exhibit a non-exponential decay, whose empirical verification would lend credence to these ideas.

\acknowledgments{}
J. F. E. C. gratefully acknowledges support from the Dodd-Walls Centre for
Photonic and Quantum Technologies.
J. L. B. acknowledges that this material is based on work supported by the
National Science Foundation under grant number 2110327.
G. Q. acknowledges funding from the FEW2MANY-SHIELD Project No. ANR-17-CE30-0015
from Agence Nationale de la Recherche.

\bibliographystyle{naturemag}
\bibliography{cite}

\begin{thebibliography}{10}
\expandafter\ifx\csname url\endcsname\relax
  \def\url#1{\texttt{#1}}\fi
\expandafter\ifx\csname urlprefix\endcsname\relax\def\urlprefix{URL }\fi
\providecommand{\bibinfo}[2]{#2}
\providecommand{\eprint}[2][]{\url{#2}}

\bibitem{danzl.haller.ea:quantum}
\bibinfo{author}{Danzl, J.~G.} \emph{et~al.}
\newblock \bibinfo{title}{Quantum gas of deeply bound ground state molecules}.
\newblock \emph{\bibinfo{journal}{Science}} \textbf{\bibinfo{volume}{321}},
  \bibinfo{pages}{1062--1066} (\bibinfo{year}{2008}).

\bibitem{ni.ospelkaus.ea:high}
\bibinfo{author}{Ni, K.-K.} \emph{et~al.}
\newblock \bibinfo{title}{A high phase-space-density gas of polar molecules}.
\newblock \emph{\bibinfo{journal}{Science}} \textbf{\bibinfo{volume}{322}},
  \bibinfo{pages}{231--235} (\bibinfo{year}{2008}).

\bibitem{croft.makrides.ea:universality}
\bibinfo{author}{Croft, J. F.~E.} \emph{et~al.}
\newblock \bibinfo{title}{Universality and chaoticity in ultracold {K+KRb}
  chemical reactions}.
\newblock \emph{\bibinfo{journal}{Nat. Commun.}} \textbf{\bibinfo{volume}{8}},
  \bibinfo{pages}{15897} (\bibinfo{year}{2017}).

\bibitem{bohn.rey.ea:cold}
\bibinfo{author}{Bohn, J.~L.}, \bibinfo{author}{Rey, A.~M.} \&
  \bibinfo{author}{Ye, J.}
\newblock \bibinfo{title}{Cold molecules: Progress in quantum engineering of
  chemistry and quantum matter}.
\newblock \emph{\bibinfo{journal}{Science}} \textbf{\bibinfo{volume}{357}},
  \bibinfo{pages}{1002--1010} (\bibinfo{year}{2017}).

\bibitem{takekoshi.reichsollner.ea:ultracold}
\bibinfo{author}{Takekoshi, T.} \emph{et~al.}
\newblock \bibinfo{title}{Ultracold dense samples of dipolar {RbCs} molecules
  in the rovibrational and hyperfine ground state}.
\newblock \emph{\bibinfo{journal}{Phys. Rev. Lett.}}
  \textbf{\bibinfo{volume}{113}}, \bibinfo{pages}{205301}
  (\bibinfo{year}{2014}).

\bibitem{park.will.ea:ultracold}
\bibinfo{author}{Park, J.~W.}, \bibinfo{author}{Will, S.~A.} \&
  \bibinfo{author}{Zwierlein, M.~W.}
\newblock \bibinfo{title}{Ultracold dipolar gas of fermionic
  $^{23}\mathrm{Na}^{40}\mathrm{K}$ molecules in their absolute ground state}.
\newblock \emph{\bibinfo{journal}{Phys. Rev. Lett.}}
  \textbf{\bibinfo{volume}{114}}, \bibinfo{pages}{205302}
  (\bibinfo{year}{2015}).

\bibitem{ye.guo.ea:collisions}
\bibinfo{author}{Ye, X.}, \bibinfo{author}{Guo, M.},
  \bibinfo{author}{Gonz{\'a}lez-Mart{\'\i}nez, M.~L.},
  \bibinfo{author}{Qu{\'e}m{\'e}ner, G.} \& \bibinfo{author}{Wang, D.}
\newblock \bibinfo{title}{Collisions of ultracold $^{23}${Na}$^{87}${Rb}
  molecules with controlled chemical reactivities}.
\newblock \emph{\bibinfo{journal}{Science Advances}}
  \textbf{\bibinfo{volume}{4}}, \bibinfo{pages}{eaaq0083}
  (\bibinfo{year}{2018}).

\bibitem{guo.ye.ea:dipolar}
\bibinfo{author}{Guo, M.} \emph{et~al.}
\newblock \bibinfo{title}{Dipolar collisions of ultracold ground-state bosonic
  molecules}.
\newblock \emph{\bibinfo{journal}{Phys. Rev. X}} \textbf{\bibinfo{volume}{8}},
  \bibinfo{pages}{041044} (\bibinfo{year}{2018}).

\bibitem{yang.zhang.ea:observation}
\bibinfo{author}{Yang, H.} \emph{et~al.}
\newblock \bibinfo{title}{Observation of magnetically tunable feshbach
  resonances in ultracold $^{23}${Na}$^{40}${K} + $^{40}${K} collisions}.
\newblock \emph{\bibinfo{journal}{Science}} \textbf{\bibinfo{volume}{363}},
  \bibinfo{pages}{261--264} (\bibinfo{year}{2019}).

\bibitem{gregory.frye.ea:sticky}
\bibinfo{author}{Gregory, P.~D.} \emph{et~al.}
\newblock \bibinfo{title}{Sticky collisions of ultracold {RbCs} molecules}.
\newblock \emph{\bibinfo{journal}{Nat. Commun.}} \textbf{\bibinfo{volume}{10}},
  \bibinfo{pages}{3104} (\bibinfo{year}{2019}).

\bibitem{gregory.blackmore.ea:loss}
\bibinfo{author}{Gregory, P.~D.}, \bibinfo{author}{Blackmore, J.~A.},
  \bibinfo{author}{Bromley, S.~L.} \& \bibinfo{author}{Cornish, S.~L.}
\newblock \bibinfo{title}{Loss of ultracold $^{87}\mathrm{Rb}^{133}\mathrm{Cs}$
  molecules via optical excitation of long-lived two-body collision complexes}.
\newblock \emph{\bibinfo{journal}{Phys. Rev. Lett.}}
  \textbf{\bibinfo{volume}{124}}, \bibinfo{pages}{163402}
  (\bibinfo{year}{2020}).

\bibitem{gersema.voges.ea:probing}
\bibinfo{author}{Gersema, P.} \emph{et~al.}
\newblock \bibinfo{title}{Probing photoinduced two-body loss of ultracold
  non-reactive bosonic $^{23}${Na}$^{87}${Rb} and $^{23}${Na}$^{39}${K}
  molecules} (\bibinfo{year}{2021}).
\newblock \eprint{2103.00510}.

\bibitem{bause.schindewolf.ea:collisions}
\bibinfo{author}{Bause, R.} \emph{et~al.}
\newblock \bibinfo{title}{Collisions of ultracold molecules in bright and dark
  optical dipole traps}.
\newblock \emph{\bibinfo{journal}{Phys. Rev. Research}}
  \textbf{\bibinfo{volume}{3}}, \bibinfo{pages}{033013} (\bibinfo{year}{2021}).

\bibitem{he.ye.ea:observation}
\bibinfo{author}{He, J.} \emph{et~al.}
\newblock \bibinfo{title}{Observation of resonant dipolar collisions in
  ultracold $^{23}\mathrm{Na}^{87}\mathrm{Rb}$ rotational mixtures}.
\newblock \emph{\bibinfo{journal}{Phys. Rev. Research}}
  \textbf{\bibinfo{volume}{3}}, \bibinfo{pages}{013016} (\bibinfo{year}{2021}).

\bibitem{nichols.liu.ea:detection}
\bibinfo{author}{Nichols, M.~A.} \emph{et~al.}
\newblock \bibinfo{title}{Detection of long-lived complexes in ultracold
  atom-molecule collisions} (\bibinfo{year}{2021}).
\newblock \eprint{2105.14960}.

\bibitem{bethe:continuum}
\bibinfo{author}{Bethe, H.~A.}
\newblock \bibinfo{title}{A continuum theory of the compound nucleus}.
\newblock \emph{\bibinfo{journal}{Phys. Rev.}} \textbf{\bibinfo{volume}{57}},
  \bibinfo{pages}{203202} (\bibinfo{year}{1940}).

\bibitem{levine:molecular}
\bibinfo{author}{Levine, R.~D.}
\newblock \emph{\bibinfo{title}{Molecular Reaction Dynamics}}
  (\bibinfo{publisher}{Cambridge University Press}, \bibinfo{year}{2005}).

\bibitem{Bohn02_PRL}
\bibinfo{author}{Bohn, J.~L.}, \bibinfo{author}{Avdeenkov, A.~V.} \&
  \bibinfo{author}{Deskevich, M.}
\newblock \bibinfo{title}{Rotational feshbach resonances in ultracold molecular
  collisions}.
\newblock \emph{\bibinfo{journal}{Phys. Rev. Lett.}}
  \textbf{\bibinfo{volume}{89}}, \bibinfo{pages}{1125--1144}
  (\bibinfo{year}{2002}).

\bibitem{mayle.ruzic.ea:statistical}
\bibinfo{author}{Mayle, M.}, \bibinfo{author}{Ruzic, B.~P.} \&
  \bibinfo{author}{Bohn, J.~L.}
\newblock \bibinfo{title}{Statistical aspects of ultracold resonant
  scattering}.
\newblock \emph{\bibinfo{journal}{Phys. Rev. A}} \textbf{\bibinfo{volume}{85}},
  \bibinfo{pages}{062712} (\bibinfo{year}{2012}).

\bibitem{mayle.quemener.ea:scattering}
\bibinfo{author}{Mayle, M.}, \bibinfo{author}{Qu\'em\'ener, G.},
  \bibinfo{author}{Ruzic, B.~P.} \& \bibinfo{author}{Bohn, J.~L.}
\newblock \bibinfo{title}{Scattering of ultracold molecules in the highly
  resonant regime}.
\newblock \emph{\bibinfo{journal}{Phys. Rev. A}} \textbf{\bibinfo{volume}{87}},
  \bibinfo{pages}{012709} (\bibinfo{year}{2013}).

\bibitem{christianen.karman.ea:quasiclassical}
\bibinfo{author}{Christianen, A.}, \bibinfo{author}{Karman, T.} \&
  \bibinfo{author}{Groenenboom, G.~C.}
\newblock \bibinfo{title}{Quasiclassical method for calculating the density of
  states of ultracold collision complexes}.
\newblock \emph{\bibinfo{journal}{Phys. Rev. A}}
  \textbf{\bibinfo{volume}{100}}, \bibinfo{pages}{032708}
  (\bibinfo{year}{2019}).

\bibitem{christianen.zwierlein.ea:photoinduced}
\bibinfo{author}{Christianen, A.}, \bibinfo{author}{Zwierlein, M.~W.},
  \bibinfo{author}{Groenenboom, G.~C.} \& \bibinfo{author}{Karman, T.}
\newblock \bibinfo{title}{Photoinduced two-body loss of ultracold molecules}.
\newblock \emph{\bibinfo{journal}{Phys. Rev. Lett.}}
  \textbf{\bibinfo{volume}{123}}, \bibinfo{pages}{123402}
  (\bibinfo{year}{2019}).

\bibitem{liu.hu.ea:photo-excitation}
\bibinfo{author}{Liu, Y.} \emph{et~al.}
\newblock \bibinfo{title}{Photo-excitation of long-lived transient
  intermediates in ultracold reactions}.
\newblock \emph{\bibinfo{journal}{Nat. Phys.}} \bibinfo{pages}{1--5}
  (\bibinfo{year}{2020}).

\bibitem{peskin.reisler.ea:on}
\bibinfo{author}{Peskin, U.}, \bibinfo{author}{Reisler, H.} \&
  \bibinfo{author}{Miller, W.~H.}
\newblock \bibinfo{title}{On the relation between unimolecular reaction rates
  and overlapping resonances}.
\newblock \emph{\bibinfo{journal}{J. Chem. Phys.}}
  \textbf{\bibinfo{volume}{101}}, \bibinfo{pages}{9672--9680}
  (\bibinfo{year}{1994}).

\bibitem{croft.bohn.ea:unified}
\bibinfo{author}{Croft, J. F.~E.}, \bibinfo{author}{Bohn, J.~L.} \&
  \bibinfo{author}{Qu\'em\'ener, G.}
\newblock \bibinfo{title}{Unified model of ultracold molecular collisions}.
\newblock \emph{\bibinfo{journal}{Phys. Rev. A}}
  \textbf{\bibinfo{volume}{102}}, \bibinfo{pages}{033306}
  (\bibinfo{year}{2020}).

\bibitem{frye.hutson:complexes}
\bibinfo{author}{Frye, M.~D.} \& \bibinfo{author}{Hutson, J.~M.}
\newblock \bibinfo{title}{Complexes formed in collisions between ultracold
  alkali-metal diatomic molecules and atoms} (\bibinfo{year}{2021}).
\newblock \eprint{2109.07435}.

\bibitem{Jachymski21_preprint}
\bibinfo{author}{Jachymski, K.}, \bibinfo{author}{Gronowski, M.} \&
  \bibinfo{author}{Tomza, M.}
\newblock \bibinfo{title}{Collisional losses of ultracold molecules due to
  intermediate complex formation.} \bibinfo{pages}{Arxiv e--prints 2110.07501}
  (\bibinfo{year}{2021}).

\bibitem{Hu_NC_13_435_2021}
\bibinfo{author}{Hu, M.-G.} \emph{et~al.}
\newblock \bibinfo{title}{Nuclear spin conservation enables state-to-state
  control of ultracold molecular reactions}.
\newblock \emph{\bibinfo{journal}{Nat. Chem.}} \textbf{\bibinfo{volume}{13}},
  \bibinfo{pages}{435} (\bibinfo{year}{2021}).

\bibitem{Quemener21_preprint}
\bibinfo{author}{Qu\'em\'ener, G.} \emph{et~al.}
\newblock \bibinfo{title}{A model for nuclear spin product-state distributions
  of ultracold chemical reactions in magnetic fields.} \bibinfo{pages}{Arxiv
  e--prints 2111.01453} (\bibinfo{year}{2021}).

\bibitem{bohr:neutron}
\bibinfo{author}{Bohr, N.}
\newblock \bibinfo{title}{Neutron capture and nuclear constitution}.
\newblock \emph{\bibinfo{journal}{Nature}} \textbf{\bibinfo{volume}{137}},
  \bibinfo{pages}{334--348} (\bibinfo{year}{1936}).

\bibitem{feshbach.weisskopf:schematic}
\bibinfo{author}{Feshbach, H.} \& \bibinfo{author}{Weisskopf, V.~F.}
\newblock \bibinfo{title}{A schematic theory of nuclear cross sections}.
\newblock \emph{\bibinfo{journal}{Phys. Rev.}} \textbf{\bibinfo{volume}{76}},
  \bibinfo{pages}{1550--1560} (\bibinfo{year}{1949}).

\bibitem{weisskopf:compound}
\bibinfo{author}{Weisskopf, V.~F.}
\newblock \bibinfo{title}{Compound nucleus and nuclear resonances}.
\newblock \emph{\bibinfo{journal}{Helvetica Physica Acta}}
  \textbf{\bibinfo{volume}{23}}, \bibinfo{pages}{187--200}
  (\bibinfo{year}{1950}).

\bibitem{feshbach.porter.ea:model}
\bibinfo{author}{Feshbach, H.}, \bibinfo{author}{Porter, C.~E.} \&
  \bibinfo{author}{Weisskopf, V.~F.}
\newblock \bibinfo{title}{Model for nuclear reactions with neutrons}.
\newblock \emph{\bibinfo{journal}{Phys. Rev.}} \textbf{\bibinfo{volume}{96}},
  \bibinfo{pages}{448--464} (\bibinfo{year}{1954}).

\bibitem{friedman.weisskopf:compound}
\bibinfo{author}{Friedman, F.~L.} \& \bibinfo{author}{Weisskopf, V.}
\newblock \bibinfo{title}{The compound nucleus}.
\newblock In \emph{\bibinfo{booktitle}{Niels Bohr and the development of
  physics}}, \bibinfo{pages}{134--162} (\bibinfo{publisher}{Pergamon},
  \bibinfo{year}{1955}).

\bibitem{feshbach:optical}
\bibinfo{author}{Feshbach, H.}
\newblock \bibinfo{title}{The optical model and its justification}.
\newblock \emph{\bibinfo{journal}{Annu. Rev. Nucl. Sci.}}
  \textbf{\bibinfo{volume}{8}}, \bibinfo{pages}{49--104}
  (\bibinfo{year}{1958}).

\bibitem{feshbach:unified}
\bibinfo{author}{Feshbach, H.}
\newblock \bibinfo{title}{Unified theory of nuclear reactions}.
\newblock \emph{\bibinfo{journal}{Ann. Phys.}} \textbf{\bibinfo{volume}{5}},
  \bibinfo{pages}{357--390} (\bibinfo{year}{1958}).

\bibitem{feshbach:unified*1}
\bibinfo{author}{Feshbach, H.}
\newblock \bibinfo{title}{A unified theory of nuclear reactions. {II}}.
\newblock \emph{\bibinfo{journal}{Ann. Phys.}} \textbf{\bibinfo{volume}{19}},
  \bibinfo{pages}{287--313} (\bibinfo{year}{1962}).

\bibitem{forst:theory}
\bibinfo{author}{Forst, W.}
\newblock \emph{\bibinfo{title}{Theory of unimolecular reactions}}
  (\bibinfo{publisher}{Elsevier}, \bibinfo{year}{2012}).

\bibitem{mitchell.richter.ea:random}
\bibinfo{author}{Mitchell, G.~E.}, \bibinfo{author}{Richter, A.} \&
  \bibinfo{author}{Weidenm\"uller, H.~A.}
\newblock \bibinfo{title}{Random matrices and chaos in nuclear physics: Nuclear
  reactions}.
\newblock \emph{\bibinfo{journal}{Rev. Mod. Phys.}}
  \textbf{\bibinfo{volume}{82}}, \bibinfo{pages}{2845--2901}
  (\bibinfo{year}{2010}).

\bibitem{christianen.groenenboom.ea:lossy}
\bibinfo{author}{Christianen, A.}, \bibinfo{author}{Groenenboom, G.~C.} \&
  \bibinfo{author}{Karman, T.}
\newblock \bibinfo{title}{Lossy quantum defect theory of ultracold molecular
  collisions.} \bibinfo{pages}{Arxiv e--prints 2108.02724}
  (\bibinfo{year}{2021}).

\bibitem{gregory.blackmore.ea:molecule-molecule}
\bibinfo{author}{Gregory, P.~D.} \emph{et~al.}
\newblock \bibinfo{title}{Molecule-molecule and atom-molecule collisions with
  ultracold rbcs molecules.} \bibinfo{pages}{Arxiv e--prints 2109.08016}
  (\bibinfo{year}{2021}).

\bibitem{hu.liu.ea:direct}
\bibinfo{author}{Hu, M.-G.} \emph{et~al.}
\newblock \bibinfo{title}{Direct observation of bimolecular reactions of
  ultracold {KRb} molecules}.
\newblock \emph{\bibinfo{journal}{Science}} \textbf{\bibinfo{volume}{366}},
  \bibinfo{pages}{1111--1115} (\bibinfo{year}{2019}).

\bibitem{Liu21_Nat}
\bibinfo{author}{Liu, Y.} \emph{et~al.}
\newblock \bibinfo{title}{Precision test of statistical dynamics with
  state-to-state ultracold chemistry}.
\newblock \emph{\bibinfo{journal}{Nature}} \textbf{\bibinfo{volume}{593}},
  \bibinfo{pages}{379} (\bibinfo{year}{2021}).

\bibitem{quemener.croft.ea:electric}
\bibinfo{author}{Qu\'em\'ener, G.}, \bibinfo{author}{Croft, J. F.~E.} \&
  \bibinfo{author}{Bohn, J.~L.}
\newblock \bibinfo{title}{Electric field dependence of complex-dominated
  ultracold molecular collisions.} \bibinfo{pages}{Arxiv e--prints 2109.02602}
  (\bibinfo{year}{2021}).

\bibitem{croft.bohn:long-lived}
\bibinfo{author}{Croft, J. F.~E.} \& \bibinfo{author}{Bohn, J.~L.}
\newblock \bibinfo{title}{Long-lived complexes and chaos in ultracold molecular
  collisions}.
\newblock \emph{\bibinfo{journal}{Phys. Rev. A}} \textbf{\bibinfo{volume}{89}},
  \bibinfo{pages}{012714} (\bibinfo{year}{2014}).

\bibitem{Klos_SR_11_10598_2021}
\bibinfo{author}{K\l{}os, J.} \emph{et~al.}
\newblock \bibinfo{title}{Roaming pathways and survival probability in
  real-time collisional dynamics of cold and controlled bialkali molecules}.
\newblock \emph{\bibinfo{journal}{Sci. Rep.}} \textbf{\bibinfo{volume}{11}},
  \bibinfo{pages}{10598} (\bibinfo{year}{2021}).

\bibitem{croft.balakrishnan.ea:long-lived}
\bibinfo{author}{Croft, J. F.~E.}, \bibinfo{author}{Balakrishnan, N.} \&
  \bibinfo{author}{Kendrick, B.~K.}
\newblock \bibinfo{title}{Long-lived complexes and signatures of chaos in
  ultracold {K}$_2$+{R}b collisions}.
\newblock \emph{\bibinfo{journal}{Phys. Rev. A}} \textbf{\bibinfo{volume}{96}},
  \bibinfo{pages}{062707} (\bibinfo{year}{2017}).

\bibitem{kendrick.li.ea:non-adiabatic}
\bibinfo{author}{Kendrick, B.~K.} \emph{et~al.}
\newblock \bibinfo{title}{Non-adiabatic quantum interference in the ultracold
  {Li} + {LiNa} $\to$ {Li}$_2$ + {Na} reaction}.
\newblock \emph{\bibinfo{journal}{Phys. Chem. Chem. Phys.}}
  \textbf{\bibinfo{volume}{23}}, \bibinfo{pages}{5096--5112}
  (\bibinfo{year}{2021}).

\bibitem{dAlessio16_PA}
\bibinfo{author}{D'Alessio, L.}, \bibinfo{author}{Kafri, Y.},
  \bibinfo{author}{Polkovnikov, A.} \& \bibinfo{author}{Rigol, M.}
\newblock \bibinfo{title}{From quantum chaos and eigenstate thermalization to
  statistical mechanics and thermodynamics}.
\newblock \emph{\bibinfo{journal}{Adv. Phys.}} \textbf{\bibinfo{volume}{65}},
  \bibinfo{pages}{239--362} (\bibinfo{year}{2016}).

\bibitem{idziaszek.julienne:universal}
\bibinfo{author}{Idziaszek, Z.} \& \bibinfo{author}{Julienne, P.~S.}
\newblock \bibinfo{title}{Universal rate constants for reactive collisions of
  ultracold molecules}.
\newblock \emph{\bibinfo{journal}{Phys. Rev. Lett.}}
  \textbf{\bibinfo{volume}{104}}, \bibinfo{pages}{113202}
  (\bibinfo{year}{2010}).

\bibitem{kotochigova:dispersion}
\bibinfo{author}{Kotochigova, S.}
\newblock \bibinfo{title}{Dispersion interactions and reactive collisions of
  ultracold polar molecules}.
\newblock \emph{\bibinfo{journal}{New J. Phys.}} \textbf{\bibinfo{volume}{12}},
  \bibinfo{pages}{073041} (\bibinfo{year}{2010}).

\bibitem{frye.julienne.ea:cold}
\bibinfo{author}{Frye, M.~D.}, \bibinfo{author}{Julienne, P.~S.} \&
  \bibinfo{author}{Hutson, J.~M.}
\newblock \bibinfo{title}{Cold atomic and molecular collisions: approaching the
  universal loss regime}.
\newblock \emph{\bibinfo{journal}{New J. Phys.}} \textbf{\bibinfo{volume}{17}},
  \bibinfo{pages}{045019} (\bibinfo{year}{2015}).

\bibitem{porter.thomas:fluctuations}
\bibinfo{author}{Porter, C.~E.} \& \bibinfo{author}{Thomas, R.~G.}
\newblock \bibinfo{title}{Fluctuations of nuclear reaction widths}.
\newblock \emph{\bibinfo{journal}{Phys. Rev.}} \textbf{\bibinfo{volume}{104}},
  \bibinfo{pages}{483--491} (\bibinfo{year}{1956}).

\bibitem{weidenmuller.mitchell:random}
\bibinfo{author}{Weidenm\"uller, H.~A.} \& \bibinfo{author}{Mitchell, G.~E.}
\newblock \bibinfo{title}{Random matrices and chaos in nuclear physics: Nuclear
  structure}.
\newblock \emph{\bibinfo{journal}{Rev. Mod. Phys.}}
  \textbf{\bibinfo{volume}{81}}, \bibinfo{pages}{539--589}
  (\bibinfo{year}{2009}).

\bibitem{bai.li.ea:model}
\bibinfo{author}{Bai, Y.-P.}, \bibinfo{author}{Li, J.-L.},
  \bibinfo{author}{Wang, G.-R.} \& \bibinfo{author}{Cong, S.-L.}
\newblock \bibinfo{title}{Model for investigating quantum reflection and
  quantum coherence in ultracold molecular collisions}.
\newblock \emph{\bibinfo{journal}{Phys. Rev. A}}
  \textbf{\bibinfo{volume}{100}}, \bibinfo{pages}{012705}
  (\bibinfo{year}{2019}).

\end{thebibliography}
\end{document}